# Photo-induced change of refractive index and transparency in $Bi_2Te_3$ films

Zengji Yue[a]*, Qinjun Chen[b], Amit Sahu[c], Xiaolin Wang[d] and Min Gu[a]

[a]Laboratory of Artificial-Intelligence Nanophotonics , School of Science, RMIT University, Melbourne, Victoria, 3001, Australia

[b]School of Physics and Electronics, Hunan University, Changsha, 410082, China

[c]Centre for Micro-Photonics, Faculty of Science, Engineering and Technology, Swinburne University of Technology, Hawthorn, Victoria 3122, Australia

[d]Institute for Superconducting and Electronic Materials, University of Wollongong, North Wollongong, NSW 2500, Australia

Email: zengji.yue@rmit.edu.au

**Abstract**

We report on an 800 nm femtosecond laser beam induced giant refractive index modulation and enhancement of near-infrared transparency in topological insulator material $Bi_2Te_3$ thin films. An ultrahigh refractive index of up to 5.9 was observed in the $Bi_2Te_3$ thin film in near-infrared frequency. The refractive index dramatically decreases by a factor of ~ 3 by an exposure to the 800 nm femtosecond laser beam. Simultaneously, the transmittance of the $Bi_2Te_3$ thin films markedly increases to ~ 96% in the near-infrared frequency. The Raman spectra provides strong evidences that the observed both refractive index modulation and transparency enhancement result from laser beam induced photooxidation effects in the $Bi_2Te_3$ thin films. The $Bi_2Te_3$ compound transfers into $Bi_2O_3$ and $TeO_2$ under the laser beam illumination. These experimental results pave the way towards the design of various optical devices, such as near-infrared flat lenses, waveguide and holograms, based on topological insulator materials.

**Introduction**

Topological insulators are quantum materials with insulating bulk states and topologically protected metallic surface states that have a Dirac dispersed band structure.[1] Fascinating electronic properties, such as quantum spin Hall effect, magnetoelectric effects, magnetic monopoles, and elusive Majorana states, are expected from topological insulators.[2] With these electronic properties, topological insulators have great potential in low loss and high efficiency spintronic and quantum computing devices.[3] Recently, various novel optical properties have also been discovered in topological insulator materials. Topological insulator material $Bi_2Se_3$ crystal displays controllable photocurrents by circularly polarized light.[4] $Bi_2Se_3$ nanosheet shows high flexibility and transparency in near-infrared ranges.[5] $Bi_2Se_3$ and $Bi_{1.5}Sb_{0.5}Te_{1.8}Se_{1.2}$ nanostructures demonstrate Dirac plasmon excitations at THz and optical frequency.[6, 7] In additions, $Bi_{1.5}Sb_{0.5}Te_{1.8}Se_{1.2}$ crystal also presents ultrahigh refractive index in the bulk.[8] With these novel optical properties, topological insulators hold great potential in design of advanced optical devices.[9, 10]

For optical devices, one of the most important parameters is the refractive index of materials. It plays a key role in almost all optical devices including lens, prism, waveguide, optical fibre, anti-reflective coating and so on. High refractive index is of great interest for imaging and lithography since the resolution scales is inversely with the refractive index.[11] High refractive-index materials have the capability to achieve broadband slow light for enhanced storage capacity of delay lines and spectral sensitivity in interferometers.[12, 13] The modulation of refractive-index enables various advanced optical devices such as ultrathin lens, multimode optical recording and digital holography.[14, 15]

Here we report on a femtosecond laser beam induced giant refractive index modulation and markedly enhancement of transparency in topological insulator material $Bi_2Te_3$ thin films grown on Si substrate. The Raman spectra show that the observed refractive index change is caused by laser beam induced photooxidation effects in the $Bi_2Te_3$ thin films. Under the laser beam illumination, the $Bi_2Te_3$ compound transfers into the compounds of $Bi_2O_3$ and $TeO_2$. These results provide a basis for designing various topological insulators based optical devices.

**Results**

Layered bismuth telluride ($Bi_2Te_3$) with a rhombohedral structure is well-known as a narrow gap semiconductor.[16] For the last decades, $Bi_2Te_3$ compound was intensively studied for high-performance thermoelectric applications at the vicinity of room temperature.[17] With angle resolved photoemission spectroscopy (ARPES), the $Bi_2Te_3$ has been verified as a typical topological insulator material.[18] In this work, the $Bi_2Te_3$ bulk crystals with a common *c*-axis orientation were grown using the Czochralski method, as described elsewhere.[19] The $Bi_2Te_3$ thin films were fabricated on Si substrate by using a pulsed laser deposition (PLD) method. The deposition substrate temperature was kept at 250°C. The chamber vacuum was $1\times10^{-4}$ Torr. The deposition time was 20 minutes. The thickness of the $Bi_2Te_3$ thin films is 40 nm.

Figure 1(a) shows the *x*-ray diffraction (XRD) patterns of topological insulator $Bi_2Te_3$ thin films grown on the Si substrate. The thin films are highly crystalline and demonstrate a preferred orientation with respect to the *c*-axis. The diffraction peaks can be indexed on the basis of the structure of rhombohedral $Bi_2Te_3$. In addition, a small (002) peak from Si substrate is visible at around 33°. Figure 1(b) shows the scanning electron microscope (SEM) of topological insulator $Bi_2Te_3$ thin films, which shows a smooth film surface.

A femtosecond-pulsed (Maitai) laser with a wavelength of 800 nm was employed for the experiments. The laser beam power was 10 mW and the spot radius was 20 μm. The laser beam repetition rate is 10 kHz and the pulse width is 100 femtoseconds. The numerical aperture of the objective is 0.7, which was used to focus the pulsed laser onto the surface of the $Bi_2Te_3$ thin film. The laser beam direct illumination was achieved by laterally translating the sample across the focal plane. All the laser radiation and treatment were accomplished in air.

Figure 1(b) displays the SEM images of $Bi_2Te_3$ thin films before laser beam illumination. Figure 1(c) shows the SEM images of the $Bi_2Te_3$ thin films after femtosecond laser beam illumination. As it can be seen, some nanoscale irregular nanoparticles were formed on the surface of $Bi_2Te_3$ thin films. This results from a rapid laser beam induced thermal annealing and oxidation on the surface of $Bi_2Te_3$ thin films under the femtosecond laser beam illumination.

The refractive index and extinction coefficient of the $Bi_2Te_3$ thin films were measured using the multiple-angle spectroscopic ellipsometer (J. A. Woollam Co. M-2000). The measured range of wavelengths was from 400 nm to 1700 nm, covering the visible and near-infrared frequency. Here, the refractive index is the effective refractive index of the $Bi_2Te_3$ thin films. Figure 2(a) shows the refractive index and extinction coefficient as a function of the wavelengths before and after femtosecond laser beam illumination. Before laser beam illumination, the refractive index monotonously increased with the wavelengths and reached up to 5.9 at the wavelength of 1700 nm. After the laser beam illumination, the refractive index still increased with the wavelengths. However, the value sharply decreased down to 2.1 at the wavelength of 1700 nm. Obviously, the femtosecond laser beam illumination induced a remarkable reduction of refractive index by a factor of ~ 3 in near-infrared range. Simultaneously, the extinction coefficient also decreased by a factor of ~ 8 at visible range.

The near-infrared transmittances of the $Bi_2Te_3$ thin films were measured using ultraviolet-visible-infrared spectroscopy. Figure 3 shows the near-infrared transmission spectra of the $Bi_2Te_3$ thin films. The transmittance markedly enhanced and reached up to 96% in the near-infrared frequency after femtosecond laser beam illumination. Recently, topological insulator $Bi_2Se_3$ nanosheets were also found to be transparent in near-infrared and could be used for transparent and flexible electrodes.[5] With laser beam induced giant refractive index modulation, topological insulators $Bi_2Te_3$ thin films are very promising candidates for near-infrared and high-performance optoelectronic devices.

**Discussion**

In order to clarify the origins of the observed optical phenomena, we conducted Raman measurements on the $Bi_2Te_3$ thin films. Figure 4 shows the Raman spectra of the $Bi_2Te_3$ thin films before and after laser beam illumination. A laser beam with 633 nm excitation was used in the experiments. Based on the Raman spectra, we confirmed that the Bi–Te thin films have $Bi_2Te_3$ stoichiometry. And the $Bi_2Te_3$ thin films were oxidized and transferred into $Bi_2O_3$ and $TeO_2$ when they were exposed to the femtosecond laser beam. The Raman peak at 126 $cm^{-1}$ was from $TeO_2$.[20] And the Raman peak at 141 $cm^{-1}$ was from α-$Bi_2O_3$.[21] Laser beam induced oxidation effects have been previously observed in $Bi_2Te_3$ nanoplates.[22] Under illumination of a 532 nm Raman laser, photo-thermal effects induced nanoparticles and photooxidation effects induced Raman peak shifts were also observed.

The refractive index of the compound $Bi_2O_3$ is below 2.5 in near-infrared range.[23, 24] Analogously, the refractive index of the compound $TeO_2$ is also below 2.5 in near-infrared range.[25] The chemical conversion from the $Bi_2Te_3$ to the $Bi_2O_3$ and the $TeO_2$ leads to the observed sharp changes of the refractive index. The giant refractive index changes enabled by photooxidation effects in $Bi_2Te_3$ thin films are orders of magnitude larger than that in the well-known photorefractive and photochromic materials.[26, 27] Based on this excellent feature, ultrathin lenses and digital hologram can be realized by controlling the nanoscale refractive index modulation using the femtosecond laser beam.

The α-$Bi_2O_3$ has a monoclinic crystal structure. It has an indirect gap of 2.6 eV and is transparent in near-infrared range.[24, 28] The $TeO_2$ crystal is commonly used in acousto-optical devices.[29] It has high infrared transparency, low light absorption and scattering.[30] The chemical conversion results in the observed large enhancements of transparency in near-infrared range. With photo-tunable refractive index and transparency, topological insulator $Bi_2Te_3$ has great potential for near-infrared transparent optical devices.

**Conclusion**

In conclusion, femtosecond laser beam induced giant changes of refractive index and transparency were observed in topological insulator material $Bi_2Te_3$ thin films. The observed changes result from the photooxidation effects when the thin films were exposed to laser beam. Based on such excellent properties, topological insulator material are promising for high-performance optoelectronic devices such as data storage and holograms. This work not only provides a deeper understanding of the interactions of light and topological insulator materials, but also paves the way towards the practical applications of topological insulator materials in optical devices.

**Figures**

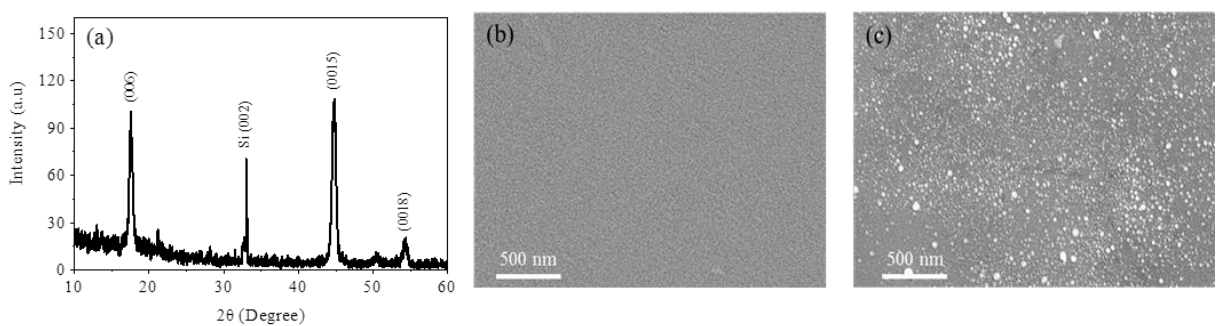

Figure 1. (a) XRD patterns of topological insulator material $Bi_2Te_3$ thin films grown on Si substrate. The thin films are highly crystalline and demonstrate a preferred orientation with respect to the *c*-axis. (b-c) SEM images of the topological insulator material Bi$_2$Te$_3$ thin films before and after femtosecond laser beam illumination. Nanoscale nanoparticles were formed after femtosecond laser beam illumination. The laser beam wavelength is 800 nm, the power is 10 mW and the spot radius is 40 μm.

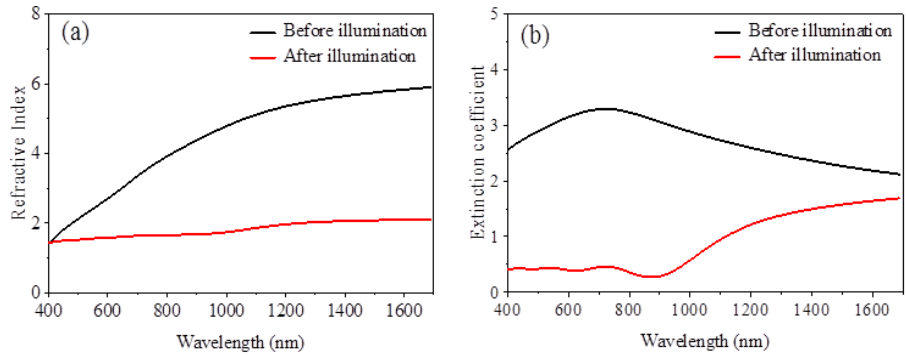

Figure 2 (a) Refractive index and (b) extinction coefficient of topological insulator material Bi$_2$Te$_3$ thin films as a function of wavelengths before and after femtosecond laser beam illumination. The refractive index and extinction coefficient were measured using the multiple-angle spectroscopic ellipsometer.

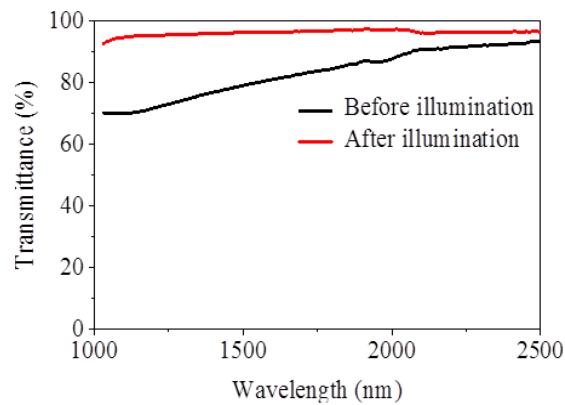

Figure 3 Near-infrared transmittance of the Bi$_2$Te$_3$ thin film before and after 800 nm femtosecond laser beam illumination. The transparency is markedly enhanced in near-infrared frequency.

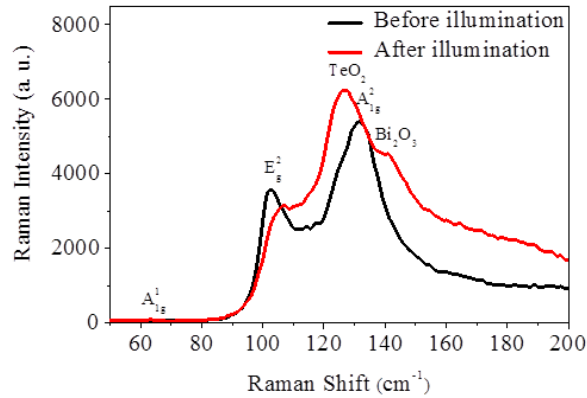

Figure 4 Raman spectra of the $Bi_2Te_3$ thin films before and after femtosecond laser beam illumination. The $Bi_2Te_3$ compound was transferred into a mixture consisting of $Bi_2O_3$ and $TeO_2$ after laser beam illumination.

**Acknowledge**

We acknowledges the support from the Australian Research Council (ARC) through the Discovery Project (DP140100849). We thank Dr. Deming Zhu and Chenglong Xu for technical assistances in Raman and UV-Vis-IR spectroscopy. This work was performed in part at the Centre for Micro-Photonics in Swinburne University of Technology.